\documentclass[seceq]{ptptex}
\usepackage{epsf}
\notypesetlogo  
\title{
\hfill{\normalsize\vbox{\hbox{\rm DPNU-03-16}  }}\\
\vspace{0.1cm}
Radiative Decays involving $a_0(980)$ and $f_0(980)$ \\
in the Vector Meson Dominance Model~\footnote{%
Talk given by M.~Harada at International Symposium on
Hadron Spectroscopy, Chiral Symmetry and 
Relativistic Description of Bound Systems, February 24-26, 2003,
Nihon University, Tokyo, Japan.}
}
\author{%
Deirdre {\sc Black}$^{\rm(a)}$,
Masayasu {\sc Harada}$^{\rm(b)}$ and Joseph {\sc Schechter}$^{\rm(c)}$
}
\inst{%
$^{\rm(a)}$ Theory Group, Jefferson Lab, 12000 Jefferson
Ave., Newport News, VA 23606, USA\\
$^{\rm(b)}$ Department of Physics, Nagoya University,
Nagoya, 464-8602, Japan\\
$^{\rm(c)}$ Department of Physics, Syracuse University
Syracuse, NY 13244-1130, USA
}
\recdate{%
\today
}
\abst{%
We summarize some features of the 
vector meson dominance model which was recently 
proposed for studying radiative decays
involving the scalar mesons.
Using the experimental values of 
$\Gamma(a_0 \rightarrow \gamma\gamma)$,
$\Gamma(f_0 \rightarrow \gamma\gamma)$ and
$\Gamma(\phi \rightarrow a_0 \gamma)$
as inputs,
we show that the model predicts a large hierarchy
between $\Gamma(a_0 \rightarrow \omega \gamma)$ and
$\Gamma(a_0 \rightarrow \rho \gamma)$ 
as well as
between $\Gamma(f_0 \rightarrow \omega \gamma)$ and
$\Gamma(f_0 \rightarrow \rho \gamma)$.
}

\begin{document}

\maketitle

\setcounter{tocdepth}{4}

\section{Introduction}

According to the recent theoretical and experimental analysis, there
is a possibility that nine light scalar mesons exist below 1 GeV, and
they form a scalar nonet~\cite{proceedings}.
In addition to the well established $f_0(980)$ and
$a_0(980)$ evidence of both experimental and theoretical 
nature for a very broad 
$\sigma$ ($\simeq560$)
and  a very broad $\kappa$ 
($\simeq900$)
has been presented.
However, the properties of the nonet members such as quark structure
and interactions with other mesons are not well known.
It is interesting to study the properties of these light scalar
mesons, which would be of great importance for a full understanding of
QCD in its nonperturbative low energy regime.

In particular, the reactions $\phi \rightarrow f_0 \gamma$ and 
$\phi \rightarrow a_0 \gamma$ have recently been observed
\cite{recentexpts} with good
accuracy and are considered as useful probes of scalar properties.
The theoretical analysis was initiated by 
Achasov and Ivanchenko~\cite{Achasov-Ivanchenko} and followed up by
many others~\cite{radiative}.
The models employed are essentially variants of the single $K$ meson
loop diagram to which a $\phi$-type vector meson, a photon
 and two pseudoscalars or a scalar
are attached.

In this paper
we summarize some features of the 
vector meson dominance model which was recently 
proposed in Ref.~\citen{BHS}
for studying radiative decays
involving the scalar mesons.
Our model predicts that
a large hierarchy
between $\Gamma(a_0 \rightarrow \omega \gamma)$ and
$\Gamma(a_0 \rightarrow \rho \gamma)$ 
as well as
between $\Gamma(f_0 \rightarrow \omega \gamma)$ and
$\Gamma(f_0 \rightarrow \rho \gamma)$.

\section{Vector Meson Dominance Model}

Framework of the model proposed in Ref.~\citen{BHS}
is that of a standard non-linear chiral Lagrangian
containing, in addition to the pseudoscalar nonet matrix field $\phi$,
the vector meson nonet matrix $\rho_\mu$ and a scalar nonet matrix
field denoted $N$.
Under chiral unitary transformations of the three light quarks;
$q_{\rm L,R} \rightarrow U_{\rm L,R} \cdot q_{\rm L,R}$, the chiral
matrix $U = \exp ( 2 i \phi/F_\pi)$,
where $F_\pi \simeq 0.131\,\mbox{GeV}$, transforms as 
$U \rightarrow U_{\rm L}\cdot U \cdot U_{\rm R}^\dagger$.
The convenient matrix
$K(U_{\rm L}, U_{\rm R}, \phi )$~\cite{CCWZ}
is defined by the following transformation property of 
$\xi$ ($U = \xi^2$):
$\xi \rightarrow U_{\rm L} \cdot \xi \cdot K^{\dag} 
  = K \cdot \xi \cdot U_{\rm R}^{\dag}$,
and specifies the transformations of ``constituent-type'' objects.
The fields we need transform as
\begin{eqnarray}
&&
  N \rightarrow K \cdot N \cdot K^{\dag} \ ,
\nonumber\\
&&
  \rho_\mu \rightarrow K \cdot \rho_\mu \cdot K^{\dag}
  + \frac{i}{\tilde{g}} K \cdot \partial_\mu K^{\dag}
\ ,
\nonumber\\
&&
  F_{\mu\nu}(\rho) = 
  \partial_\mu \rho_\nu - \partial_\nu \rho_\mu - i 
  \tilde{g} \left[ \rho_\mu \,,\, \rho_\nu \right]
  \rightarrow
  K \cdot F_{\mu\nu} \cdot K^{\dag}
\ ,
\label{transf}
\end{eqnarray}
where the coupling constant $\tilde{g}$ is about $4.04$.
One may refer to Ref.~\citen{Harada-Schechter} for our treatment of
the 
pseudoscalar-vector Lagrangian and to
Ref.~\citen{Black-Fariborz-Sannino-Schechter:99} for the scalar
addition.
The entire Lagrangian is chiral invariant (modulo the quark mass term
induced symmetry breaking pieces) and, when
electromagnetism is added, gauge invariant.

\begin{figure}[htbp]
\begin{center}
\epsfxsize = 6.5cm
\ \epsfbox{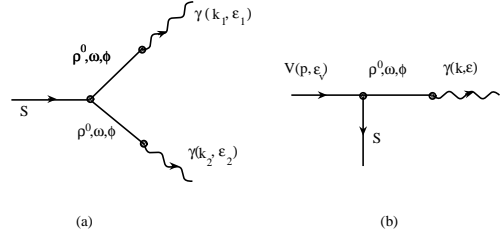}
\end{center}
\caption[]{%
Feynman diagrams for (a)~$S\rightarrow \gamma\gamma$ and
(b)~$V \rightarrow S \gamma$.
}\label{fig:1}
\end{figure}

In Ref.~\citen{BHS}, 
the strong trilinear scalar-vector-vector terms were included
into the effective Lagrangian as
\begin{eqnarray}
&&
{\cal L}_{SVV} =  \beta_A \,
\epsilon_{abc} \epsilon^{a'b'c'}
\left[ F_{\mu\nu}(\rho) \right]_{a'}^a
\left[ F_{\mu\nu}(\rho) \right]_{b'}^b N_{c'}^c
\nonumber\\
&& \quad
{}+
 \beta_B \, \mbox{Tr} \left[ N \right]
\mbox{Tr} \left[ F_{\mu\nu}(\rho) F_{\mu\nu}(\rho) \right]
\nonumber\\
&& \quad
{}+
 \beta_C \, \mbox{Tr} \left[ N F_{\mu\nu}(\rho) \right]
\mbox{Tr} \left[ F_{\mu\nu}(\rho) \right]
\nonumber\\
&& \quad
{}+
 \beta_D \, \mbox{Tr} \left[ N \right]
\mbox{Tr} \left[ F_{\mu\nu}(\rho) \right]
\mbox{Tr} \left[ F_{\mu\nu}(\rho) \right]
\ .
\label{SVV}
\end{eqnarray}
Chiral invariance is evident from (\ref{transf}) and the four
flavor-invariants are needed for generality.  (A term 
$\sim \mbox{Tr}( FFN )$ is linearly dependent on the four shown).
Actually the $\beta_D$ term will not contribute in our model so there
are only three relevant parameters $\beta_A$, $\beta_B$ and $\beta_C$.
Equation~(\ref{SVV}) is analogous to the $PVV$ 
interaction~\footnote{%
 It was shown~\cite{BKY:88-HY:03} that the complete vector
 meson dominance (VMD) is violated in the 
 $\omega\rightarrow \pi^0 \pi^+ \pi^-$ decay which
 is expressed
 by $PVV$ interactions.  However, since the VMD is satisfied
 in other processes such $\pi^0 \rightarrow \gamma \gamma^\ast$
 as well as
 in the electromagnetic form factor of pion, we here assume that
 it holds in the processes related to $SVV$ interactions.
}
which was
originally introduced as a $\pi\rho\omega$ coupling a long time 
ago~\cite{GSW}. 
One can now compute the amplitudes for 
$S\rightarrow\gamma\gamma$ and $V \rightarrow S \gamma$ according to
the diagrams of Fig.~\ref{fig:1}.

The decay matrix element for $S \rightarrow \gamma \gamma$ is
written as 
$(e^2/\tilde{g}^2) X_S \times \bigl(
k_1\cdot k_2 \, \epsilon_1\cdot \epsilon_2 -
k_1\cdot \epsilon_2 \, k_2 \cdot \epsilon_1
\bigr)$ 
where $\epsilon_\mu$ stands for the photon polarization vector.  It is
related to the width by
\begin{equation}
\Gamma \left( S \rightarrow \gamma\gamma \right)
=
\alpha^2 \frac{\pi}{4} m_S^3
\left\vert \frac{X_S}{\tilde{g}^2} \right\vert^2
\ ,
\label{swidth}
\end{equation}
and $X_S$ takes on the specific forms:
\begin{eqnarray}
X_{\sigma} &=&
  \frac{4}{9} \beta_A 
    \left( \sqrt{2} s - 4 c \right)
  {} + \frac{8}{3} \beta_B
    \left( c - \sqrt{2} s \right)
\ ,
\nonumber\\
X_{f_0} &=&
  - \frac{4}{9} \beta_A 
    \left( \sqrt{2} c + 4 s \right)
  {} + \frac{8}{3} \beta_B
    \left( \sqrt{2} c + s \right)
\ , 
\nonumber\\
X_{a_0} &=& \frac{4\sqrt{2}}{3} \beta_A 
\ .
\label{samps}
\end{eqnarray}
Here $\alpha=e^2/(4\pi)$, $s = \sin \theta_S$ and $c = \cos\theta_S$
where 
the scalar
mixing angle, $\theta_S$ is defined from~\cite{BHS}
\begin{equation}
\left(\begin{array}{c}
\sigma \\ f_0
\end{array}\right)
=
\left(\begin{array}{cc}
c & -s \\ s & c
\end{array}\right)
\left(\begin{array}{c}
N_3^3 \\ (N_1^1 + N_2^2)/\sqrt{2}
\end{array}\right)
\ .
\label{mixing}
\end{equation}
Furthermore ideal mixing for the vectors, with
$\rho^0 = (\rho_1^1 - \rho_2^2)/\sqrt{2}$,
$\omega = (\rho_1^1 + \rho_2^2)/\sqrt{2}$,
$\phi = \rho_3^3$, was assumed for simplicity.

Similarly, the decay matrix element for $V\rightarrow S \gamma$ is
written as $(e/\tilde{g}) C_V^S \times \left[
p\cdot k \epsilon_V \cdot \epsilon 
- p \cdot \epsilon k \cdot \epsilon_V
\right]$.
It is related to the width by
\begin{equation}
\Gamma (V \rightarrow S \gamma ) =
\frac{\alpha}{3} \left\vert k_V^S \right\vert^3
\left\vert \frac{C_V^S}{\tilde{g}} \right\vert^2
\ ,
\end{equation}
where $k_V^S = (m_V^2 - m_S^2)/(2m_V)$ is the photon momentum in the
$V$ rest frame.
For the energetically allowed $V \rightarrow S \gamma$ processes we
have
\begin{eqnarray}
C_\phi^{f_0} &=& 
 \frac{2\sqrt{2}}{3} \beta_A c
 - \frac{4}{3} \beta_B \left( \sqrt{2} c + s \right)
 + \frac{\sqrt{2}}{3} \beta_C
   \left( c - \sqrt{2} s \right)
\ , 
\nonumber\\
C_\phi^{\sigma} &=& 
 - \frac{2\sqrt{2}}{3} \beta_A s
 - \frac{4}{3} \beta_B 
   \left( c - \sqrt{2} s \right)
 - \frac{2}{3} \beta_C
   \left( c + \frac{1}{\sqrt{2}} s \right)
\ , 
\nonumber\\
C_\phi^{a_0} &=& \sqrt{2} \left( \beta_C - 2 \beta_A \right)\ , 
\nonumber\\
C_\omega^{\sigma} &=& 
 \frac{2\sqrt{2}}{3} \beta_A 
   \left( c + \sqrt{2} s \right)
 + \frac{2\sqrt{2}}{3} \beta_B 
   \left( c - \sqrt{2} s \right)
 - \frac{2}{3} \beta_C
   \left( \sqrt{2} c + s \right),
\nonumber\\
C_{\rho^0}^\sigma &=& -2\sqrt{2} \beta_A c + 2 \sqrt{2} \beta_B
\left( c- \sqrt{2} s \right).
\label{vamps}
\end{eqnarray}

In addition, the same model predicts amplitudes for
 the energetically allowed
$S\rightarrow V \gamma$ processes:
$f_0 \rightarrow \omega \gamma$, $f_0 \rightarrow \rho^0 \gamma$,
$a_0^0 \rightarrow \omega \gamma$,
$a_0^0 \rightarrow \rho^0 \gamma$ and,
if $\kappa^0$ is sufficiently heavy 
$\kappa^0 \rightarrow K^{\ast0} \gamma$.
The corresponding width is
\begin{equation}
\Gamma (S \rightarrow V \gamma ) =
 \alpha \left\vert k_S^V \right\vert^3
\left\vert \frac{D_S^V}{\tilde{g}} \right\vert^2
\ ,
\end{equation}
where $k_S^V = (m_S^2 - m_V^2)/(2m_S)$ and
\begin{eqnarray}
D_{f_0}^\omega &=& 
 \frac{2}{3} \beta_A \left( - 2 c +\sqrt{2} s \right)
 + \frac{2}{3} \beta_B \left( 2 c +\sqrt{2} s \right)
 + \frac{2}{3} \beta_C
   \left( c - \sqrt{2} s \right)
\ , 
\nonumber\\
D_{f_0}^{\rho^0} &=& 
 - 2 \sqrt{2} \beta_A s
 + 2 \beta_B 
   \left(2 c +\sqrt{2} s \right)
\ , 
\nonumber\\
D_{a_0}^\omega &=& 2 \beta_C \ ,
\nonumber\\
D_{a_0}^{\rho^0} &=&  \frac{4}{3} \beta_A \ ,
\nonumber\\
D_{\kappa^0}^{K^{\ast0}} &=& 
 - \frac{8}{3} \beta_A 
\ .
\label{svgamps}
\end{eqnarray}

\section{Results}

Let us summarize main points of the results obtained in 
Ref.~\citen{BHS} together with new results from our recent
analysis~\cite{BHS:prep}.

We should stress again that
all the different decay amplitudes are described by the parameters
$\beta_A$, $\beta_B$ and $\beta_C$.
Below we shall first
illustrate the procedure for the model of a single
putative scalar nonet~\cite{Black-Fariborz-Sannino-Schechter:99}.

We determine the value of $\beta_A$ 
from the $a_0 \rightarrow \gamma\gamma$ process.
Substituting 
$\Gamma_{\rm exp}(a_0\rightarrow \gamma\gamma) =
(0.28\pm0.09)\,\mbox{keV}$ (obtained using \cite{PDG}
$B(a_0 \rightarrow K \bar{K})/B( a_0 \rightarrow \eta \pi) =
0.177 \pm 0.024$)
into Eqs.~(\ref{swidth}) and (\ref{samps})
yields
\begin{equation}
\beta_A = (0.72\pm0.12)\,\mbox{GeV} \ ,
\label{valA}
\end{equation}
where we assumed positive in sign.
By using this value, 
the value of $\beta_C$ is determined 
from 
$\Gamma_{\rm exp}(\phi \rightarrow a_0 \gamma) =
(0.47\pm0.07)\,\mbox{keV}$ 
(obtained by assuming $\phi\rightarrow
\eta\pi^0\gamma$ is dominated by $\phi\rightarrow a_0\gamma$)
and Eq.~(\ref{vamps}) 
as
\begin{equation}
\beta_C = 
(7.7\pm0.5 \,,\, -4.8\pm0.5)\,\mbox{GeV}^{-1} \ .
\label{valC}
\end{equation}
We stress that 
the values of $\beta_A$ and $\beta_C$ obtained above 
are independent of the mixing angle $\theta_S$.
It should be noticed that $\vert\beta_A\vert$ is almost an order of
magnitude smaller than $\vert\beta_C\vert$.
As we can see from Eq.~(\ref{svgamps}),
the amplitude $D_{a_0}^\omega$ is given by $\beta_C$ while
$D_{a_0}^{\rho^0}$ is given by only $\beta_A$.
Then, the large hierarchy between $\beta_C$ and $\beta_A$ implies
that there is a large hierarchy between 
$\Gamma(a_0\rightarrow\omega\gamma)$ and 
$\Gamma(a_0\rightarrow\rho\gamma)$.
Actually, by using the values of $\beta_A$ and $\beta_C$ given
in Eqs.~(\ref{valA}) and (\ref{valC}),
they are estimated as
\begin{eqnarray}
&& \Gamma(a_0\rightarrow\omega\gamma) 
  = \left( 641\pm87\,,\,251\pm54 \right) \,\mbox{keV} \ ,
\nonumber\\
&& \Gamma(a_0\rightarrow\rho\gamma) 
  =  3.0\pm1.0 \,\mbox{keV} \ .
\end{eqnarray}
This implies that there is a large hierarchy between
$\Gamma(a_0\rightarrow\omega\gamma)$ and
$\Gamma(a_0\rightarrow\rho\gamma)$
which is caused by an order of magnitude difference between
$\vert\beta_C\vert$ and $\vert\beta_A\vert$.

We next determine the value of $\beta_B$ from 
the $f_0\rightarrow \gamma\gamma$ process.
$X_{f_0}$ in Eq.~(\ref{samps}) depends on
$\beta_B$ as well as on $\beta_A$ and the mixing angle $\theta_S$.
Here we take the mixing angle as 
\begin{equation}
\theta_S \simeq - 20^{\circ}\ ,
\end{equation}
which is characteristic of $qq{\bar q}{\bar q}$ type 
scalars~\cite{Black-Fariborz-Sannino-Schechter:99}.
By using this and the value of $\beta_A$ in Eq.~(\ref{valA}),
$\Gamma_{\rm exp}(f_0 \rightarrow \gamma\gamma) = 
0.39\pm0.13\,\mbox{keV}$ yields
\begin{equation}
\beta_B = (0.61\pm0.10 \,,\,-0.62\pm0.10)\,\mbox{GeV}^{-1} \ .
\end{equation}
This implies that $\vert\beta_B\vert$ is on the order of 
$\vert\beta_A\vert$, and almost an order of magnitude smaller
than $\vert\beta_C\vert$.
Equation~(\ref{svgamps}) shows that
$D_{f_0}^\omega$ includes $\beta_C$ while
$D_{f_0}^\rho$ does not.
Thus, we have a large hierarchy between 
decay widths of
$f_0\rightarrow\omega\gamma$ and
$f_0\rightarrow\rho\gamma$:
The typical predictions are given by
\begin{eqnarray}
&& \Gamma(f_0\rightarrow\omega\gamma) = ( 88\pm17)\,\mbox{keV}
\ ,
\nonumber\\
&& \Gamma(f_0\rightarrow\rho\gamma) = ( 3.3\pm2.0)\,\mbox{keV}
\ .
\label{pred f0}
\end{eqnarray}
This implies that there is a large hierarchy between
$\Gamma(f_0\rightarrow\omega\gamma)$ and
$\Gamma(f_0\rightarrow\rho\gamma)$ which is 
caused by the fact that
$\vert\beta_C\vert$ is 
an order of magnitude larger than
$\vert\beta_A\vert$ and $\vert\beta_B\vert$.

Let us check the dependence of the above results on the choice
of the scalar mixing angle $\theta_S$.
In Ref.~\citen{Black-Fariborz-Sannino-Schechter:99},
the value of $\theta_S \simeq -90^{\circ}$ was obtained as
another solution to reproduce the masses of the 
lightest scalar nonet,
although the predicted value of $f_0$-$\pi$-$\pi$ coupling
is much larger than the value obtained in 
Ref.~\citen{Harada-Sannino-Schechter}
by fitting to
the $\pi\pi$ scattering amplitude 

As we stressed above, the values of $\beta_A$ and $\beta_C$ are
independent of the scalar mixing angle $\theta_S$.
The value of $\beta_B$ determined from 
$\Gamma(f_0\rightarrow\gamma\gamma)$ becomes
\begin{equation}
\beta_B = 
(1.1\pm0.1 \,,\, 0.12\pm0.13)\,\mbox{GeV}^{-1} \ .
\label{valB90}
\end{equation}
Then the typical predictions for $\Gamma(f_0\rightarrow\omega\gamma)$
and $\Gamma(f_0\rightarrow\rho\gamma)$ are given by
\begin{eqnarray}
&& \Gamma(f_0\rightarrow\omega\gamma) = ( 86\pm16)\,\mbox{keV}
\ ,
\nonumber\\
&& \Gamma(f_0\rightarrow\rho\gamma) = ( 3.4\pm3.2)\,\mbox{keV}
\ .
\end{eqnarray}
These predictions are very close to the ones in 
Eq.~(\ref{pred f0}).
This can be understood by the following consideration:
{}From the expression of $D_{f_0}^{\omega}$ in 
Eq.~(\ref{svgamps}),
we can see that it is 
dominated by the term including $\beta_C$
which is proportional to
$(\cos\theta_S - \sqrt{2} \sin\theta_S)$.
Then, the approximate relation
\begin{equation}
\cos( - 20^{\circ} ) - \sqrt{2} \, \sin( - 20^{\circ} )
\simeq
\cos( - 90^{\circ} ) - \sqrt{2} \, \sin( - 90^{\circ} )
\simeq 1.4 
\label{eq20-90}
\end{equation}
implies that
the value of $D_{f_0}^{\omega}$ for $\theta_S=-90^{\circ}$
is close to that for $\theta_S=-20^{\circ}$, and thus
$\Gamma(f_0\rightarrow\omega\gamma)$ for $\theta_S=-90^{\circ}$
to that for $\theta_S=-20^{\circ}$.
As for $\Gamma(f_0\rightarrow\rho\gamma)$ we note that
the following relation is satisfied for
$X_{f_0}$ in Eq.~(\ref{samps}) 
and $D_{f_0}^{\rho^0}$ in Eq.~(\ref{svgamps}):
\begin{equation}
3 X_{f_0} - 2 \sqrt{2} D_{f_0}^{\rho^0}
= - \frac{4}{3} \sqrt{2} \beta_A ( c - \sqrt{2} s )
\ .
\end{equation}
Since we use the experimental value of 
$\Gamma(f_0\rightarrow\gamma\gamma)$, i.e., $X_{f_0}$ as
an input, this relation implies that
the predicted value of $\Gamma(f_0\rightarrow\rho\gamma)$
for $\theta_S=-90^{\circ}$ is roughly equal
to that for $\theta_S=-20^{\circ}$.
{}From the above consideration we conclude that
there is a large hierarchy between 
$\Gamma(f_0\rightarrow\omega\gamma)$
and $\Gamma(f_0\rightarrow\rho\gamma)$
for $\theta_S=-20^{\circ}$ and $\theta_S=-90^{\circ}$:
\begin{equation}
\Gamma(f_0\rightarrow\omega\gamma)
\gg
\Gamma(f_0\rightarrow\rho\gamma)
\quad \mbox{for} \
\theta_S=-20^{\circ} \ \mbox{and}\ \theta_S=-90^{\circ}
\ .
\end{equation}

\section{Discussion}

In this paper we showed the predictions of our model
only for $\Gamma(a_0\rightarrow\omega\gamma)$,
$\Gamma(a_0\rightarrow\rho\gamma)$,
$\Gamma(f_0\rightarrow\omega\gamma)$
and $\Gamma(f_0\rightarrow\rho\gamma)$.
Predictions on several other processes such as 
$\Gamma(\sigma\rightarrow \gamma\gamma)$ 
and $\Gamma(\phi \rightarrow \sigma \gamma)$
can be seen in Ref.~\citen{BHS}.

The value of $\Gamma(\phi\rightarrow f_0 \gamma)$ predicted
in Ref.~\citen{BHS} is considerably smaller than the
experimental value
of $\Gamma(\phi \rightarrow f_0 \gamma)$.
We may need to include the effect of $K$-loop
which would give a large enhancement as shown
in Ref.~\citen{Achasov-Ivanchenko}.
We leave the analysis with $K$-loop corrections to
future publications~\cite{BHS:prep}.

\section*{Acknowledgements}

We are happy to thank N. N. Achasov for a suggested correction
to the first version of Ref.~\citen{BHS} and with
A. Abdel-Rahiem and A. H. Fariborz for very
helpful discussions.
D.B. wishes to acknowledge support from the Thomas Jefferson National
Accelerator Facility operated by the Southeastern Universities Research
Association (SURA) under DOE contract number DE-AC05-84ER40150.
The
work of J.S. is supported in part by DOE contract DE-FG-02-85ER40231.

\end{document}